\documentclass[%
 aip,
 jcp,%
 amsmath,amssymb,
preprint,%
]{revtex4-1}

\usepackage{graphicx}
\usepackage{dcolumn}
\usepackage{bm}
\usepackage{verbatim}

\draft

\begin{document}

\title{Volume Phase Transition of Polyelectrolyte Gels: Effects of Ionic Size}

\author{Li-Jian~Qu}
\affiliation{Department of Physics, Beijing Normal University, Beijing 100875}
\author{Xinghua~Zhang}
\email{zhangxh@bjtu.edu.cn}
\affiliation{School of Science, Beijing Jiaotong University, Beijing 100044, China}
\author{Jiuzhou~Tang}
\affiliation{Beijing National Laboratory for Molecular Sciences, Institute of Chemistry, Chinese Academy of Sciences, Beijing 100190, China}
\author{Dadong~Yan}
\email{yandd@bnu.edu.cn}
\affiliation{Department of Physics, Beijing Normal University, Beijing 100875}

\date{\today}

\begin{abstract}
Although the volume transition of the polyelectrolyte gel has been studied for decades, little research on the effects of size of the mobile ions has been conducted. In the present paper, Tanaka's classical theory of polyelectrolyte gel is extended to the cases of mobile ions of finite volume. In the salt free limit, the theoretical results show that the discontinuous volume transition of the polyelectrolyte gel will become a continuous one with an increase of the counterionic size. An increase in salt concentration can also make the polyelectrolyte gel in poor solvent collapse. Poorer solvent is needed to trigger the salt-induced collapse in polyelectrolyte gel with larger mobile ions than that with smaller ones. The effects of ionic size on the critical points and phase diagram of the volume transition are also discussed. The theoretical results suggest that the swelling behavior of polyelectrolyte gel might be tuned with salt of different volumes.
\end{abstract}

\maketitle

\section{Introduction}

Polyelectrolyte gels, cross-linked networks formed by polyelectrolyte chains, can undergo tremendous reversible volume change in response to the changes of solvent quality, temperature, added salt, pH, and external field, etc. \cite{tanaka1980phase,shibayama1993volume,khokhlov1993conformational,quesada2011gel} This distinctive character leads to the wide applications of the polyelectrolyte gel in various technologies, such as in drug delivery,\cite{kabanov2008nanogels,kabanov2009nanogels} disposable diapers,\cite{odio2000diaper} microsensors,\cite{richter2008review} adaptive liquid microlenses,\cite{dong2006adaptive} medicine.\cite{lee2009thermally} Therefore, volume transition of polyelectrolyte gel is still being widely studied today, \cite{schneider2004discontinuous,bodrova2007influence,yin2009swelling,hua2012theory,jha2011electrostatic,jha2012understanding} more than three decades after Tanaka's pioneering work.\cite{tanaka1980phase}

Tanaka\cite{tanaka1980phase,ricka1984swelling} first observed the discontinuous volume transition of polyelectrolyte gel in experiment and explained it by incorporating the osmotic pressure of the counterions into the Flory-Rehner's theory for neutral polymer gel .\cite{flory1943I,flory1943II} They observed a discontinuous swollen-collapsed transition in polyelectrolyte gels due to the osmotic pressure of the counterions. Since then the theory of polyelectrolyte gel's volume transition has been an active field, and the Flory-Rehner-Tanaka theory formalism has been extended to many directions. Tanaka soon incorporated the salt effects into his theory.\cite{ohmine1982salt} Khokhlov and Kramarenko took the possibility of ion-pair into account.\cite{khokhlov1994polyelectrolyte} Safronov et al. considered the ion type dependence of the Flory-Huggins parameter.\cite{safronov2009influence} English et al. studied the volume transition of polyampholyte gels using Tanaka's theoretical formalism.\cite{english1996equilibrium} Even though some more sophisticated theories have been developed in recent years, \cite{yin2009swelling,longo2011molecular,hua2012theory} Tanaka's formalism is still valuable in understanding polyelectrolyte gel's static behaviors for its simplicity in both concept and mathematics.

Most of the previous studies took the mobile ions as point charge, and few studies considered the finite size of the mobile ions. However, significant impacts of ionic size have been observed in many polyelectrolyte systems. Cuvillier et al. found that the steric repulsion had great effects on adsorption of large ions onto a charged monolayer. \cite{cuvillier1997adsorption} Their experiment was soon explained by a modified Poisson-Boltzmann theory taking into account the ion volume. \cite{borukhov2000adsorption} The same theory formalism was also applied to examine the ion effects of charge renormalization of cylindrical and spherical macroions. \cite{borukhov2004charge} Ahrens et al. found a nonlinear osmotic regime for polyelectrolyte brush and they explained this new regime with a scaling theory taking the counterionic size into account.\cite{ahrens2004nonlinear} So there is a reason to believe that it is valuable to explore the effects of ionic size in polyelectrolyte gel systems.

Recently, Jha and collaborators theoretically explored the effects of ionic size and dielectric mismatch on volume transition of polyelectrolyte nanogels in good solvent. \cite{jha2012understanding}. They predicted that finite volume of ions would make nanogels undergo a re-entrant transitions with increase in the salt concentration. Bodrova and Potemkin examined the effects of counterion's size on polyelectrolyte gels.\cite{bodrova2007influence} They predicted continuous swollen-collapsed transitions in polyelectrolyte gels with counterions large enough. Besides the volume of the counterion, they also examined the effects of short-range interaction of counterion-counterion and counterion-polymer monomer. However, they did not study the effects of added salt which is a common experiment practice. They mainly focused on the swelling behavior of the polyelectrolyte gel and paid little attention to the phase diagram of the volume transition. Up to now, there still have been few experimental studies exploring the effect of ion volume.

Our aim in this study is to include the finite size of the ions in Tanaka's theoretical formalism for polyelectrolyte gel and study how the phase diagram of the volume transition is affected. Both cases of salt free and salt added are studied. The dielectric mismatch between the gel and solvent is not taken into account at the present stage.

\section{Model and Theory}

The swelling behaviors of a polyelectrolyte gel is described by the Tanaka theory,\cite{ricka1984swelling} which combined the Flory-Rehner theory\cite{flory1943I,flory1943II} for the neutral polymer gel and Donnan theory for mobile ions. This classical formalism is still widely used to study the polyelectrolyte gel due to its simplicity and extensibility, as reviewed by Manuel Quesada-Perez et al. \cite{quesada2011gel} We consider a polyelectrolyte gel with a volume fraction of $\phi_0$ at the reference state. All subchains of the gel are identical and are composed of $N$ segments. The subchains are negatively charged with a dissociation degree $\alpha$. The polyelectrolyte gel is immersed into an aqueous solution containing monovalent salt. When the solvent quality and salt concentration are changed, a volume transition occurs to the polyelectrolyte gel. The volume fraction becomes $\phi$. We assume that this is a homogeneous deformation. For simplicity, the volume of a polymer segment and a solvent molecule are equal to $v_0$, and all the mobile ions are of the same volume of $v_{ion}=u v_0$. In the following discussion, we use $v_0$ as the volume unit and $k_BT$ as the energy unit, $k_B$ and $T$, respectively, as the Boltzmann constant and absolute temperature. The dimensionless number density of mobile ions in the gel are denoted as $n_+$ and $n_-$ for positive and negative ions respectively. The number density of the salt in the bulk solution is $n_s$.

The free energy per polymer segment can be written as

\begin{equation}\label{eqn:freeg}
    F = F_{el} + f/\phi
\end{equation}
The first term is the elastic free energy, adopting the following Flory-Rehner's form
\begin{equation}\label{eqn:FRelastic}
    F_{el} = \frac{3}{2N}\left [\left( \frac{\phi_0}{\phi} \right)^{2/3}-\frac{1}{3}\ln \left ( \frac{\phi_0}{\phi} \right )\right ]
\end{equation}
$f$ in the second term is a sum of two contributions: the translational entropy of the small mobile ions, $f_{ion}$, and the Flory-Huggins mixing energy, $f_{mix}$. The translational entropy of mobile ions can be expressed as:
\begin{equation}\label{eqn:ionentropy}
    f_{ion} =  n_+\left[\ln\left (\frac{n_+}{n_s}\right)-1\right] + n_-\left[\ln\left(\frac{n_-}{n_s}\right)-1\right] + 2 n_s + \psi_D (n_+ - n_- - \alpha\phi)
\end{equation}
where $\psi_D$ is electrostatic Donnan potential, a constant to ensure the local electroneutrality. The Flory-Huggins mixing free energy can be presented as
\begin{equation}\label{eqn:mix}
    f_{mix} = [1-\phi-u(n_++n_-)]\left[\ln\left(\frac{1-\phi-u(n_++n_-)}{1-2un_s}\right)-1\right]+1-2un_s+\chi\phi(1-\phi)
\end{equation}

The equilibrium state is determined by minimizing $F$,
\begin{equation}
    \frac{\partial f}{\partial n_+} = \frac{\partial f}{\partial n_-} = 0
\end{equation}
\begin{equation}
    \frac{\partial F}{\partial \phi} = 0
\end{equation}
Then we obtain the equations for the equilibrium ion number density and polymer volume fraction,
\begin{equation}\label{eqn:iondens}
    \frac{n_+n_-}{[1-\phi-u(n_++n_-)]^{2u}} = \frac{n_s^2}{(1-2un_s)^{2u}}
\end{equation}
\begin{equation}\label{eqn:LEA}
    n_+ = n_- + \alpha\phi
\end{equation}
\begin{equation}\label{eqn:chiphi}
    \frac{\phi_0}{N} \left [ \left( \frac{\phi}{\phi_0} \right)^{1/3}-\frac{\phi}{2\phi_0} \right ] + \ln\left[\frac{1-\phi-u(n_++n_-)}{1-2un_s}\right] +\chi \phi^2  + \phi + u (n_+ + n_- - 2n_s) = n_+ + n_- - 2n_s
\end{equation}
The right side of Eq.~\ref{eqn:chiphi} is the osmotic pressure of mobile ions. The excluded volume effect of ions is included in the terms containing $u$ in the left side of Eq.~\ref{eqn:chiphi}.
\section{Results and Discussions}
\subsection{Polyelectrolyte gel with salt free}
We start with the salt free limit, i.e., $n_s=0$ and $n_-=0$. From Eqs.~\ref{eqn:iondens},~\ref{eqn:LEA} and~\ref{eqn:chiphi}, the equilibrium polymer volume fraction satisfies the following equation,
\begin{equation}\label{eqn:chisaltfree}
    \frac{\phi_0}{N} \left [ \left( \frac{\phi}{\phi_0} \right)^{1/3}-\frac{\phi}{2\phi_0} \right ] + \ln[1-(1+\alpha u)\phi] +\chi \phi^2 + (1+\alpha u)\phi = \alpha\phi
\end{equation}
If we take $u = 0$, Eq.~\ref{eqn:chisaltfree} reduces to the prediction of point counterion.\cite{tanaka1980phase} 

\begin{figure}
  \centering
  \includegraphics[scale=0.3]{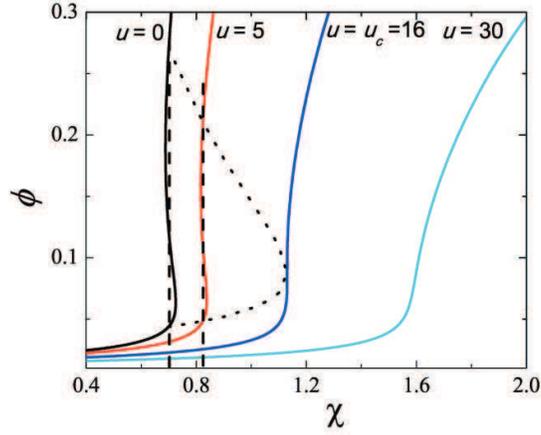}\\
  \caption{Swelling curve of polyelectrolyte gel without added salt. The dissociation degree is $\alpha = 0.02$. The length of the subchain is $N = 50$, and the volume fraction at the reference state is $\phi_0 = 0.05$. $u = u_c = 16$ is the critical value of counterion volume. The dashed lines on curves of $u = 0 $ and $u = 5$ indicate the position of the discontinuous volume transition. The dotted line indicates the coexistence line of the swollen and the collapsed state for the polyelectrolyte gel.} \label{fgr:collpsaltfree}
\end{figure}

Fig.~\ref{fgr:collpsaltfree} shows the solvent quality dependence of polymer concentration for polyelectrolyte gel with $\phi_0 = 0.05$, $N = 50$ and $\alpha = 0.02$ for different values of conterion volume. 
In non-poor solvent($\chi \leq 0.5$), the size of counterion has insignificant effect on the volume of the swollen polyelectrolyte gel. As $\chi$ increases, the counterionic size begin to show influences gradually. For polyelectrolyte gels dissociating large counterions ($u \geq 16$ in Fig.~\ref{fgr:collpsaltfree}), their volume decrease continuously with the decrease of the solvent quality. This is reflected in that $\chi(\phi)$ is a monotonous function. The $\chi(\phi)$ is a non-monotonous function for small $u$ ($u \leq 16$ in Fig.~\ref{fgr:collpsaltfree}). In a range of $\chi$, three values of the polymer volume fraction $\phi$ correspond to the same $\chi$. The lowest and the highest of the three values of $\phi$ correspond to minima of the free energy. When the two values of $\phi$ correspond to the same free energy, the polyelectrolyte will undergo a first order volume transition at the corresponding $\chi$. At the transition point, the gel may be in a coexistence state of collapsed and swollen conformation. The coexistence curve is presented in Fig.~\ref{fgr:collpsaltfree} as dotted lines, which are obtained by equalizing the free energies of the collapsed and swollen state. For the volume transition of the polyelectrolyte gel, there is a critical value $u_c$ for the counterion's volume, which can be determined according to
\begin{equation}\label{eqn:criticalcondition}
   \frac{\partial^{3} F}{\partial \phi^{3}} = 0
\end{equation}
Incorporating with Eq.~\ref{eqn:chisaltfree}, we obtain the critical value $u_c = 16$. For smaller counterions, $~u< u_c$, the gel will undergo a jumpwise volume transition with the increase of $\chi$. The positions of the first-order phase transition are marked with dashed lines in Fig.~\ref{fgr:collpsaltfree}. For larger counterions, $u \geq u_c$, the $\phi(\chi)$ exhibits a monotonous dependence on $\chi$, indicating that the polyelectrolyte gel dissociating larger ions collapse continuously with the decrease of the solvent quality.

\begin{figure}
  \centering
  \includegraphics[scale=0.3]{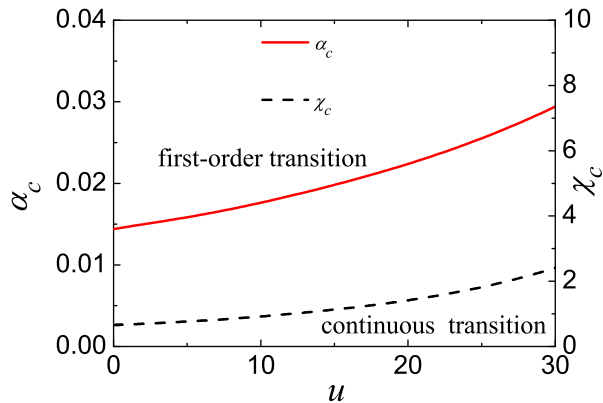}\\
  \caption{Critical lines for polyelectrolyte gel with salt free. The length of the subchain is $N=50$, and the volume fraction at the reference state is $\phi_0=0.05$. }\label{fgr:criticalsaltfree}
\end{figure}

In Fig.~\ref{fgr:criticalsaltfree}, the critical lines are shown for polyelectrolyte gel with $\phi_0=0.05$ and $N=50$ without salt. Both the critical values of charge fraction $\alpha_c$ (solid line) and Flory-Huggins $\chi_c$ (dash line) parameter are presented. Under the lines, the volume of the polyelectrolyte gel will undergo a continuous volume transition with the decrease of the solvent quality (increase of $\chi$). On the contrary, when $\alpha>\alpha_c$ ($\chi>\chi_c$), a first-order volume transition takes place. The critical lines terminate at the upper limit of the dissociation degree, imposed by $\alpha \leq 1$ and $\alpha \leq (1-\phi)/(u\phi)$, of the polyelectrolyte gel (not shown in Fig.~\ref{fgr:criticalsaltfree}). Beyond this ending point, the polyelectrolyte gel can only experience a continuous volume transition with the variation of the solvent quality. The figure shows that the critical values increase with the ion volume. This indicates that we need a poorer solvent to make the polyelectrolyte gel collapsed. The reason is that the volume of the mobile ions enhances both the effective excluded volume (the two-body interaction) and three-body interaction between the polymer segments. This can be seen clearly as follows. For small values of $(1+\alpha u)\phi$, the logarithm in Eq.~\ref{eqn:chisaltfree} can be expanded as $\ln[1-(1+\alpha u)\phi]\approx -(1+\alpha u)\phi -[-(1+\alpha u)\phi]^2/2+[-(1+\alpha u)\phi]^3/3\ldots$. Then the effective excluded volume is $v=(1+\alpha u)^2-2\chi$ and the three-body interaction coefficient is $w=(1+\alpha u)^3/6$, both of which are enhanced by the ion volume. Consequently, poorer solvent is needed to cause the polyelectrolyte gel to collapse. When the volume of the counterion is large enough, the critical values $\chi_c$ are so high that no first-order transition occurs for polyelectrolyte gel in common poor solvent. We can also understand the continuous volume transition in polyelectrolyte gel dissociating large counterions. In this case, the effective excluded volume interaction, estimated as $[(1+\alpha u)^2/2-\chi]\phi^2$, dominates over the osmotic pressure of the counterions, $\alpha \phi$, making the polyelectrolyte gel behaves like neutral polymer gel. The results obtained here, taking the translational entropy of all the mobile ions and solvent molecules into account, is consistent with the previous work of Bodrova and Potemkin, \cite{bodrova2007influence} where the virial approximation was adopted to compute the free energy of excluded volume of ions.
\subsection{Polyelectrolyte gel with added salt}

\begin{figure}
\centering
\includegraphics[scale=0.25]{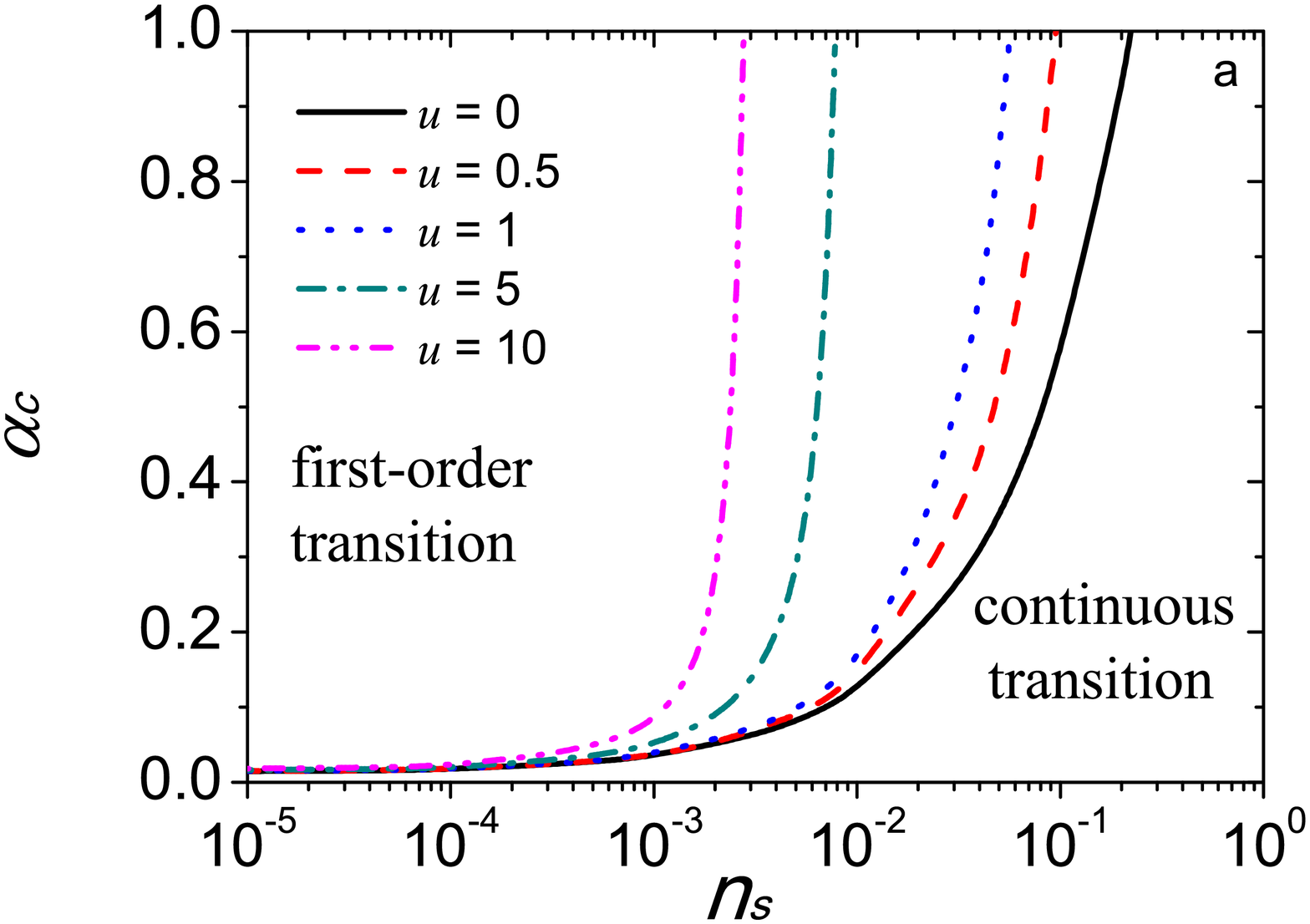}
\includegraphics[scale=0.25]{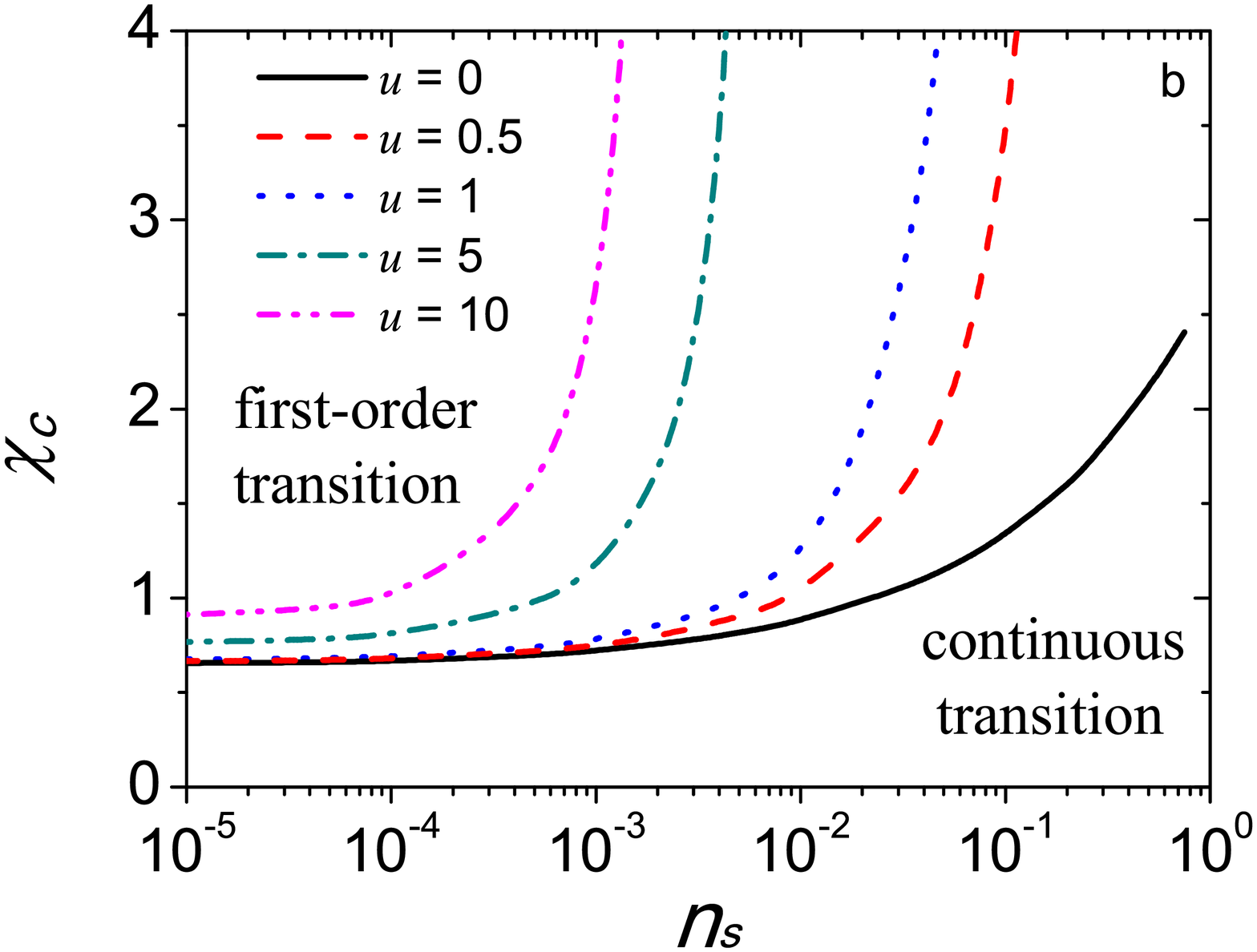}
\caption{Critical line for polyelectrolyte gel: (a) dissociation degree, (b) Flory-Huggins parameter. The length of the subchain is $N=50$, and the volume fraction at the reference state is $\phi_0=0.05$.}
\label{fgr:criticalsalt}
\end{figure}

Now we turn to the volume transition of polyelectrolyte gel with added salt. The equilibrium ion density and polymer volume fraction satisfy Eqs.~\ref{eqn:iondens},~\ref{eqn:LEA} and~\ref{eqn:chiphi}. Similar to case of salt free, the volume transition induced by decrease of solvent quality may be a continuous or a jumpwise one. We first discuss the the boundary between the continuous and the first-order transition, that is the critical line. The salt dependence of the critical lines of $\alpha_c(n_s)$ and $\chi_c(n_s)$ for ions of various sizes, which are determined by using Eq.~\ref{eqn:criticalcondition}, are presented in Figs.~\ref{fgr:criticalsalt}a~and~\ref{fgr:criticalsalt}b  respectively. When the salt concentration concentration is low enough, the critical values are almost constant and equal to that of salt free case, as shown in Fig.~\ref{fgr:criticalsaltfree}. In this situation, the salt concentration is too low to show any influence. However, even at low salt concentration situation, poorer solvent is needed, as shown clearly in Fig.~\ref{fgr:criticalsalt}b, to make the first-order collapse occur for larger mobile ions. With the increase of the salt concentration, the salt dependence of the critical values appears gradually. This dependence turns up much earlier for larger mobile ions. Moreover, the salt dependence is more significant for larger ions. For larger ions, the critical value nearly jumps to quite a high value at a certain salt concentration. So at higher concentration of larger mobile ions, no discontinuous volume transition can be observed. The disappearance of the first-order transition is due to that the osmotic pressure of the ions are far outweighed by the effective excluded volume interaction that are greatly enhanced by larger ions. The polyelectrolyte gel immersed in a solution of large ions is similar to neutral polymer gel, and collapses continuously with the decrease of the solvent quality. For smaller ions, the salt concentration just raises the threshold for the first-order volume transition.

\begin{figure}
  \centering
  \includegraphics[scale=0.3]{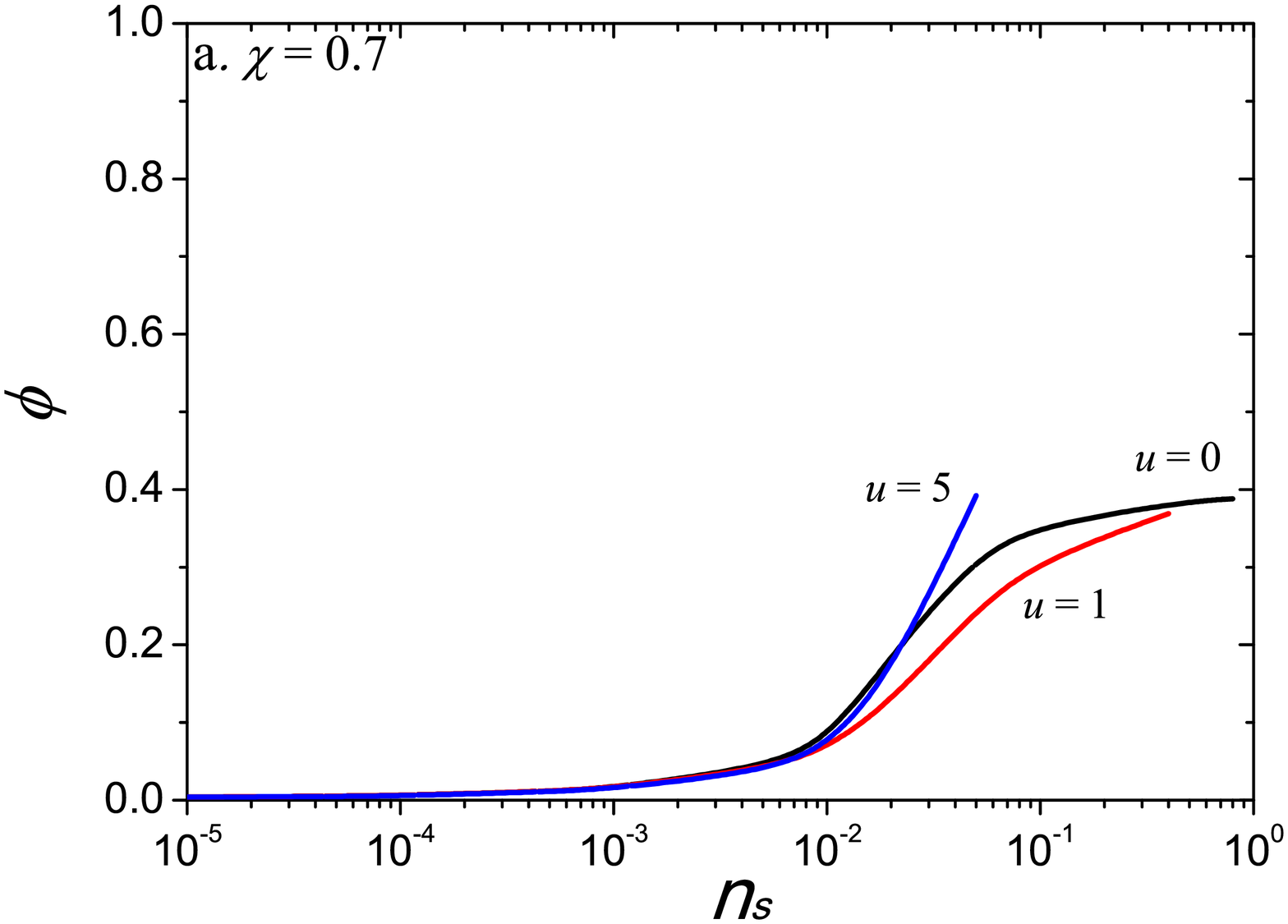}
  \includegraphics[scale=0.3]{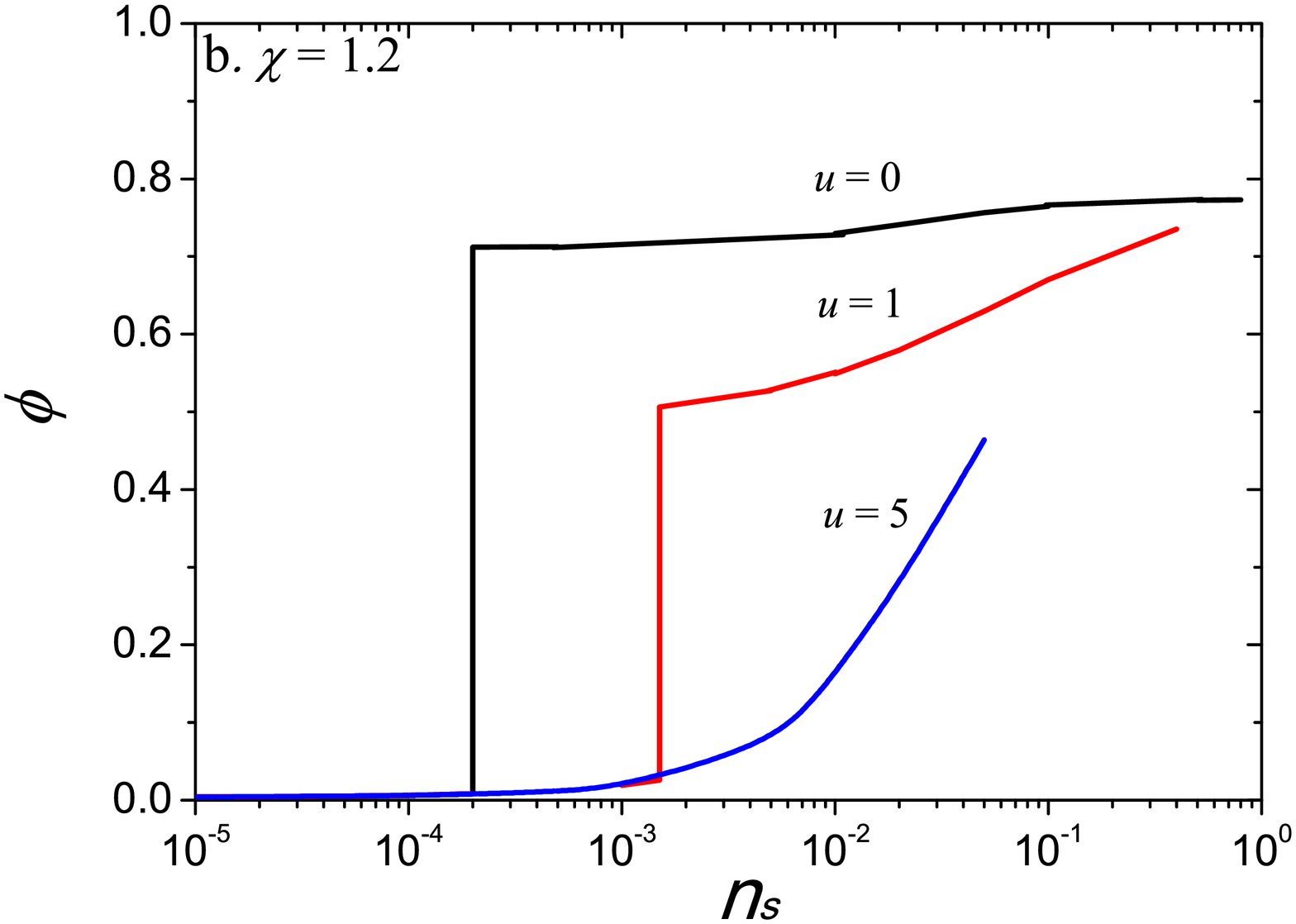}
  \includegraphics[scale=0.3]{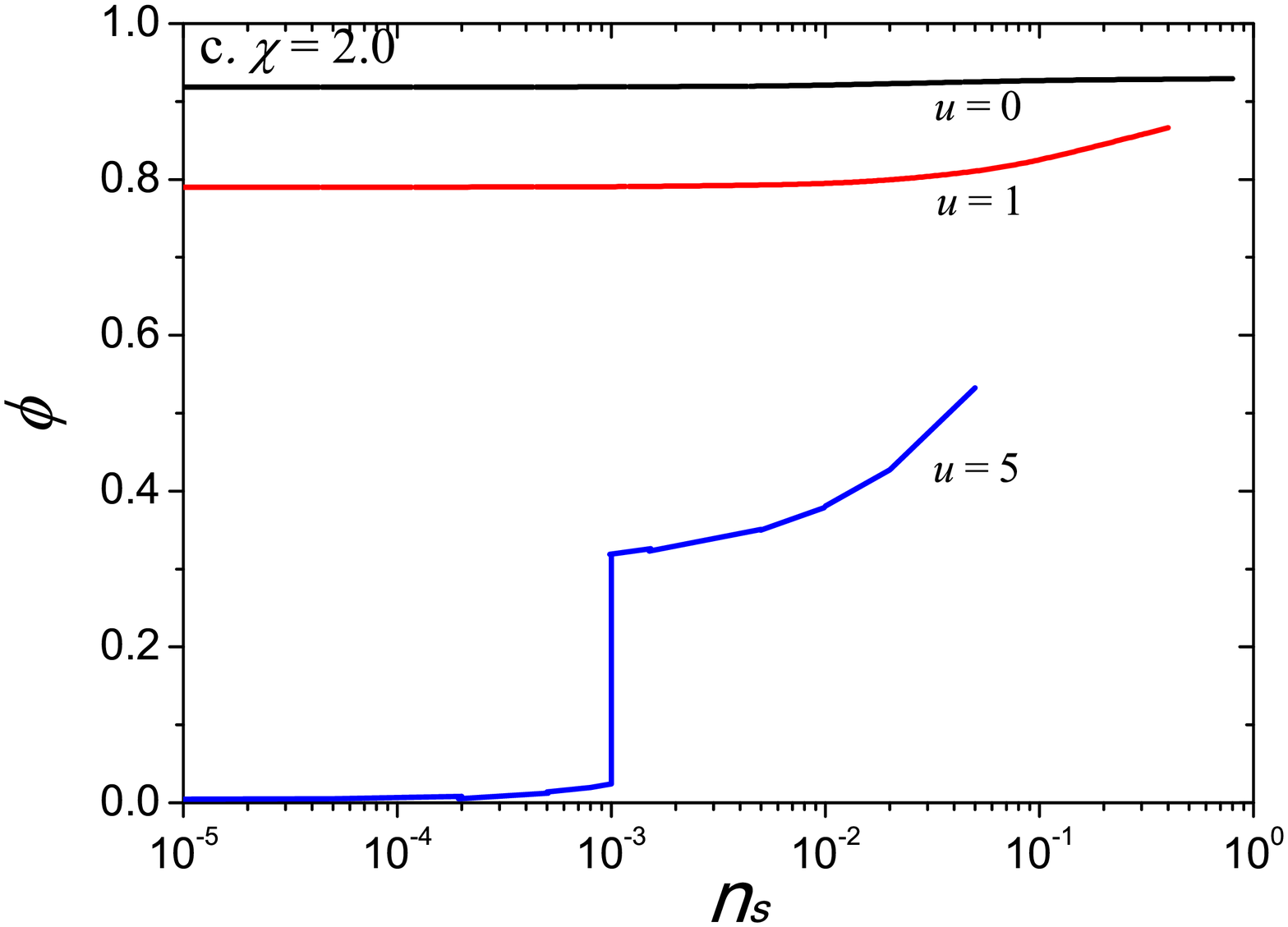}
  \caption{Salt concentration dependence of polymer concentration for salt of different sizes $u = 0$,~$1$ and~$5$ and solvents of different quality: (a) $\chi = 0.7$, (b) $\chi = 1.2$ and (c) $\chi = 2.0$. The dissociation degree is $\alpha=0.02$. The length of the subchain is $N=50$, and the volume fraction at the reference state is $\phi_0=0.05$.}\label{fgr:saltcollapse}
\end{figure}

For neutral polymer gel, the collapse transition can be only driven by the decrease of solvent quality (or temperature in experiment practice). However, the collapse of polyelectrolyte gel in poor solvent can also be caused by adding salt. For solvents of different quality, the effect of the concentration of salt has different behaviors, as shown in Fig.~\ref{fgr:saltcollapse}, in which $\chi=0.7$, $1.2$ and~$2.0$, respectively. Near the $\theta$ point (i.e. $\chi=0.7$), the volume of the polyelectrolyte gel decreases continuously with the increase of the salt concentration. The solvent is not poor enough to make the gel collapse. When the salt concentation is low, the osmotic pressure of the mobile ions swells the gel. As the salt concentration increases, the osmotic pressure lowers gradually, resulting the contraction of the polyelectrolyte gel. For poorer solvent, $\chi=1.2$, one can see that a discontinuous collapse transition for gel along with point ions or small ions when the salt concentration exceeds a certain value. The ion's volume delays the discontinuous collapse transition. For larger ions, $u=5$, the polyelectrolyte gel still undergoes a continuous contraction, since $\chi=1.2<\chi_c$ in this case. For even poorer solvent, $\chi=2.0$, the salt concentration can trigger a sudden collapse for the polyelectrolyte gel along with large ions. In such a poor solvent, polyelectrolyte gels in solution of small ions are already collapsed. Consequently, the increase of salt concentration causes only a slight decrease in the gel volume.

\begin{figure}
  \centering
  \includegraphics[scale=0.3]{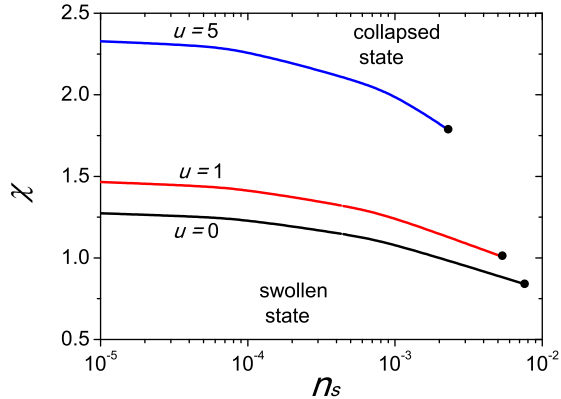}\\
  \caption{Coexistence line of the swollen and collapsed polyelectrolyte gel in coordinates Flory-Huggins parameter $\chi$ vs salt concentration $n_s$. The dissociation degree is $\alpha=0.02$. The length of the subchain is $N=50$, and the volume fraction at the reference state is $\phi_0=0.05$. The black point on each line denotes the critical point at which the coexistence line ends.}\label{fgr:saltbinodal}
\end{figure}

In Fig.~\ref{fgr:saltbinodal}, the coexistence lines of the collapsed and swollen state for polyelectrolyte gel with dissociation degree $\alpha = 0.02$ are plotted. The ion volume is labelled to the corresponding line. The gel is in swollen state under the corresponding line and collapsed state above the line. The lines terminate at the critical points. Beyond the critical point, one cannot distinguish the swollen and the collapsed state. When we tune the solvent quality and the salt concentration, the polyelectrolyte gel will undergo a first order transition when crossing the coexistence line. We can see clearly from the phase diagram that the ion's volume has great influence on the volume transition of the polyelectrolyte gel. Poorer solvent is needed to make first order transition happen for larger ions when the added salt are of the same concentration. This can be attributed to enhancement of the effective excluded volume interaction because of the ion volume, as we discuss the polyelectrolyte gel with salt free.

\section{Conclusion}

This paper presented the influences of ionic size on the volume transition of the polyelectrolyte gel, based on Tanaka-Flory-Rehner-Donnan theory. In both cases of salt free and salt added, conditions of discontinuous volume transition are more difficult to reach for polyelectrolyte gels with larger ions. In the salt free limit, the polyelectrolyte gel will collapse smoothly with the decrease of the solvent quality when the dissociated counterions are large enough. The reason is that the effective excluded volume interaction between the polymer segments enhanced by the counterionic size significantly surpass osmotic pressure of the mobile ions. When salt of large ions is added, a continuous volume transition of the polyelectrolyte gel will be observed at high salt concentration. At low salt concentration, much poorer solvent is needed to trigger the first-order volume transition for polyelectrolyte gel with larger ions than that with smaller ions. The conditions of the coexistence of the swollen and collapse state are also predicted. However, the present theory cannot capture the chain conformation, and more advanced theory or simulation is needed to study the coexistent state.

The theoretical predictions of the present study is testable in experiments. Sufficiently large ions can be obtained by dissolving organic salt in the solution. So the swelling behavior of polyelectrolyte gel can also be tuned by adding salt of different volumes.

The ionic size dependence of the polyelectrolyte gel behaviors can be taken as one of the influences of ion specificity. So other aspects of ion specificity may also play a significant role in polyelectrolyte gel behaviors. It is necessary to systematically explore ionic specific effects on polyelectrolyte gel.
\\

\noindent
\textbf{ACKNOWLEDGMENTS}

The authors acknowledge the financial support of National Natural Science Foundation of China (NSFC) 20990234, 973 Program of the Ministry of Science and Technology (MOST) 2011CB808502, and the Fundamental Research Funds for the Central Universities No. 2013JBM087.

\bibliography{VPTJCP}

\end{document}